\documentstyle[twocolumn,fleqn]{article}

\begin{document}
\onecolumn
\title{
{\Large Gravitational action versus entropy on simplicial lattices in
four dimensions\thanks{Supported in part by "{}Fonds zur F\"orderung der
wissenschaftlichen Forschung".} \thanks{To be published in the Proceedings of
Lattice 92, Amsterdam, The Netherlands, 1992, eds. J.~Smith and P.~van
Baal, Nucl.~Phys.~B (Proc.~Suppl.)}}}
\author{
W.~Beirl, E.~Gerstenmayer, H.~Markum and J.~Riedler \\[0.5cm]
Institut f\"ur Kernphysik, Technische Universit\"at Wien,
A-1040 Vienna, Austria}

\maketitle

\begin{abstract}
We investigate quantum gravity on simplicial lattices using Regge
calculus with special emphasize on the problem of the unbounded action.
The role of the entropy for the path integral is discussed in detail. Our
numerical results show further evidence for the existence of an entropy
dominated region with well defined expectation values even for unbounded
action. Analyses are performed both for the standard regular triangulation of
the 4-torus and for irregularly triangulated lattices
obtained by insertion of vertices using barycentric subdivision.
\end{abstract}

\twocolumn

\section{INTRODUCTION}
Quantum Theory and General Relativity are, separately, the two most
successful theories of $20^{th}$ century physics. Unfortunately, their
combination to a theory of Quantum Gravity is plagued by serious problems.
A quantum field theory based on the Einstein-Hilbert action (here in
Euclidean form with the Planck length $L_P$, the metric $g$ and the
curvature scalar $R$)
\begin{equation}
-I_E = L_P^{-2} \int d^4x g^{\frac{1}{2}}R
\end{equation}
is perturbatively not renormalizeable and thus
requires non-perturbative methods such as lattice field theory
techniques. A natural starting point for investigations of this kind is
the sum-over-histories formulation that leads to the Euclidean
path integral \cite{hartle}
\begin{equation}
Z = \int D{\bf g}e^{-I_E({\bf g})}.
\end{equation}
The integration extends over a certain class of closed 4-geometries {\bf g}
each weighted with the corresponding gravitational action.

To get a computationally manageable theory one needs to specify the
ingredients of the above formula. The integration that should
include a summation over different topologies has to be restricted to a
classifiable set of 4-geometries \cite{hartle}.
For the measure $D{\bf g}$ different reasonings lead to different
functional forms \cite{menotti}. This ambiguity makes independent
investigations of the various proposals necessary.
The action $I_E({\bf g})$ suffers from
divergencies due to fluctuations of the conformal factor. However, this
might be no problem if the entropy of the system suppresses geometries
with large action \cite{berg}.

An elegant method to evaluate the path integral (2) approximately
is provided by the
Regge calculus (Sec.~2). It is well suited to study the question
of entropy which is discussed in Sec.~3, but as usual in lattice theories
the dependence on the underlying lattice structure has to be clarified
(Sec.~4). The presentation of our results is followed
by a conclusion in Sec.~5.

\section{REGGE CALCULUS}

The main idea of the Regge calculus is to approximate 4-geometries by
simplicial lattices, i.e.~piecewise flat geometries constructed
from 4-simplices glued together at their surfaces \cite{hartle,regge}.
A simplicial lattice is characterized by its incidence matrices
specifying how the simplices are contained in each other and by the
squared lengths $q_l$ of the links $l$. For Euclidean
configurations all generalized triangle inequalities have to be
fulfilled.
The Regge action is given by
\begin{equation}
-I_R = L^{-2}_P \sum_t 2A_t \delta_t,
\end{equation}
where the summation extends over all triangles $t$ with area $A_t$
and deficit angle
\begin{equation}
\delta_t=2\pi - \sum_{s\supset t}\Theta_{s,t}.
\end{equation}
In four dimensions the deficit angle denotes the difference between
$2\pi$ and the sum of the dihedral angles $\Theta_{s,t}$ of all
4-simplices $s$ sharing $t$.

On lattices with finite (squared) fatness
\begin{equation}
\phi_s = C^2 \frac{{V_s}^2}{max_{l\subset s}(q_l^4)} \geq f > 0
\end{equation}
the Regge action corresponds to the Einstein-Hilbert action in the
classical continuum limit \cite{classcont}.
For each 4-simplex $s$ with volume $V_s$ we introduce a cutoff $f$ for the
fatness and set the constant $C=24$.

In our simulations we use the path integral
\begin{equation}
Z = \int \prod_ldq_l{\cal F}_f(q_l)e^{\beta\sum_tA_t\delta_t
- \lambda\sum_sV_s}.
\end{equation}
The uniform measure $\prod_ldq_l$ is the simplest choice within the Regge
calculus but we expect universality of the results as suggested
by previous investigations \cite{beirl1}. The function
${\cal F}_f(q_l)$ restricts the integration to Euclidean geometries
with finite fatness. It is zero if either the generalized triangle
inequalities or the condition (5) is violated and has the value one otherwise.
We express all quantities in units of $L_P$ and set the scale by the
coupling parameter $\beta$. The cosmological constant $\lambda$ is
introduced to fix the average volume of the lattice and set to one,
$\lambda=1$.

The observables we will discuss are the average curvature in units of
the average link length (with number of links $N_1$),
\begin{equation}
R =  \frac{ \sum_t A_t \delta_t } { \sum_s V_s }
\left( \frac{1}{N_1} \sum_l q_l \right),
\end{equation}
and the average squared deficit angle $\delta_t^2$.
While the expectation value $\langle R \rangle$ clearly represents an
interesting quantity the importance of $\langle \delta_t^2 \rangle$
can be seen from the following simple argument.

Introducing an additional term $-\alpha \sum_t \delta_t^2$ in the exponent
of Eq.~(6) one gets for the derivatives of $\ln Z$
\begin{eqnarray*}
\frac{\partial \ln Z}{\partial \beta}=
    \langle \sum_t A_t \delta_t \rangle     ,&&
\frac{\partial \ln Z}{\partial \alpha} =
    - \langle \sum_t \delta_t^2 \rangle
\end{eqnarray*}
\begin{eqnarray}
     \Longrightarrow \hspace{1cm}
\frac{\partial \langle \delta_t^2 \rangle}{\partial \beta} &=&
  - \frac{\partial \langle A_t \delta_t \rangle}{\partial \alpha}.
\end{eqnarray}
We perform our computations for $\alpha = 0$ but expect that a slight
increase of $\alpha$ would lead to decreasing absolute values of the average
action since flat lattices are prefered. We conclude that
$\langle \delta_t^2 \rangle$ decreases with $\beta$ if
the average curvature $\langle R \rangle$ is negative.
Since previous computations indicate a negative curvature in the entropy
dominated region \cite{berg,beirl1,beirl2,hamber} one expects
an interesting behavior of $\langle \delta_t^2 \rangle$.

\section{ENTROPY}

As mentioned before the unbounded action is one of the problems of the
Euclidean quantum-gravity path-integral persisting also in its simplicial
form. It was recognized first by Berg \cite{berg} that the entropy of the
system could suppress configurations with large curvature. The path
integral
\begin{equation}
Z = \int_{-\infty}^{+\infty}dIn(I)e^{- \beta I}
\end{equation}
exists if the state density $n(I)$ associated with the entropy of the
system vanishes fast enough for $I \to -\infty$. A state
density of the form $n(I) = e^{-\beta_c |I|}$ for example would lead to a path
integral with finite radius $| \beta| < \beta_c$ of
convergence.

We use Monte Carlo methods to investigate this possibility.
This does not give the form of $n(I)$ directly but
the stability of the system can be studied.
Introducing a lower limit $f>0$ for the
fatness $\phi_s$ of each 4-simplex $s$ the action is bounded
since the divergent modes of the system leading to very long and thin
simplices are cut off. Decreasing $f$ systematically one has to
analyze the convergence of the expectation values
to find out if an entropy dominated phase exists.

Results from this procedure are plotted in Fig.~1 and 2 showing
the average curvature $\langle R \rangle$ and
the average squared deficit angle $\langle \delta_t^2 \rangle$, respectively.
Calculations are performed with a regular triangulation of the 4-torus
obtained from a hypercubic tesselation \cite{berg,beirl1,beirl2,hamber} and
averages are taken over at least 2500 sweeps for each data point after
thermalization.
The existence of a phase with small and negative $\langle R \rangle$
for all values of $f$ is seen clearly.
While in the range $10^{-3} \geq f > 10^{-5}$ the shape of the curves and
their zeros change sig\-nificantly,
%
%
one finds for $f \leq 10^{-5}$ that the results agree within error bars.
This suggests that configurations with large curvature and very small
fatness do not contribute to the path integral but are suppressed
due to their small entropy factor.

\begin{figure}[t]
\begin{center}
\input{fig1R}
\end{center}
\caption{Average curvature $\langle R \rangle$ as a function of
$\beta$ on a regularly
triangulated 4-torus with $4^4$ vertices for decreasing cutoff $f$.
Error bars due to mean standard deviation are smaller
than the symbols, dotted lines are guides for the eye.
For all values of $f$ the entropy dominated phase is seen.
No significant cutoff dependence is found for $f \leq 10^{-5}$.}
\end{figure}
\begin{figure}[t]
\begin{center}
\input{fig1d}
\end{center}
\caption{Average squared deficit angle $\langle \delta_t^2 \rangle$
versus $\beta$ for the same configurations as in Fig.~1.
In the entropy dominated region $\langle \delta_t^2 \rangle$ decreases
with $\beta$ reaching a minimum value near $\langle R \rangle = 0$.}
\end{figure}
\begin{figure}[t]
\begin{center}
\input{fig1h}
\end{center}
\caption{History of the average curvature $R$ from homogeneous (lower
curve) and
inhomogeneous (upper curve) start configurations.
No cutoff is imposed ($f=0$) and the
coupling ($\beta=0.03$) lies in the entropy dominated region. One time unit
$\tau$ corresponds to 10 sweeps. After about 10k sweeps the two curves
converge to the negative expectation value of $R$.}
\end{figure}
The behavior of $\langle \delta_t^2 \rangle$ (Fig.~2) is in
accordance with argument (8) of the previous Section.
In the entropy dominated phase one finds a decreasing
$\langle \delta_t^2 \rangle$ with increasing $\beta$.
The minimum values of $\langle \delta_t^2 \rangle$ correspond to
$\langle R \rangle = 0$.
Since Eq.~(8) repre\-sents a relation between thermodynamical expectation
values this is a certain indication for the equilibrium of the phase
with negative curvature.

Studying the convergence of $R$ for different start configurations
should give further hints for the existence of a well defined phase.
We concentrate on the most interesting case without cutoff, $f=0$.
Fig.~3 shows the history of $R$ at a coupling $\beta = 0.03$ in the well
defined region for a homogeneous start configuration and for an
inhomogeneous start configuration with positive curvature.
One observes that configurations with positive curvature are not favored.
After about 10k sweeps the value of $R$ converges towards the
corresponding negative expectation value. For larger $\beta$ or for start
configurations with larger $R$ long runs are needed to see the convergence.
In general we expect a finite range $0 \leq \beta < \beta_c$ where the
average curvature $\langle R \rangle$ is small independent of the start
configuration. From our data we extrapolate on the $4^4$-lattice for
$f \to 0$ a value of $\beta_c \approx 0.065$.
\begin{figure}[t]
\begin{center}
\input{fig2R}
\end{center}
\caption{Average curvature $\langle R \rangle$ as a function of $\beta$
for $f = 5 \times 10^{-4}$ on regular and irregular triangulations of
the 4-torus with increasing number $B$ of inserted vertices. The values of
$\langle R \rangle$ are shifted towards positive values for $B>0$.}
\end{figure}

\section{LATTICE STRUCTURE}

The dependence of the results on the underlying triangulation is an open
but crucial question within Regge calculus. A simple way to con\-struct
irregularly triangulated lattices is to start with the hypercubic
triangulation and to insert vertices using barycentric subdivision
\cite{beirl2}. A new vertex of this kind is
shared by 5 4-simplices, 10 tetrahedrons, 10 triangles and 5 links.
Different to the regular triangulation each of the new triangles is
shared by three 4-simplices only.
One therefore expects positive deficit angles at
these triangles leading to a growth of $\langle R \rangle$.

In Fig.~4 the average curvature $\langle R \rangle$ for lattices with
$B = 0, 5, 10$
and $20$ inserted vertices and $f = 5 \times 10^{-4}$ is depicted.
We observe for increasing $B$ a shift towards positive values
of $\langle R \rangle$.
Detailed investigations of the
configurations show that the additional vertices tend to develop spikes while
the others behave like vertices of a regular lattice \cite{beirl2}.
Decreasing $f$ allows larger spikes at the additional vertices. This
leads to longer autocorrelation lengths making
relyable investigations of the limit $f \to 0$ very
time consuming.

With a slight variation of the above construction, which has 5 links per
new vertex, one can insert vertices with larger coordination numbers.
Preliminary results suggest that the tendency to form spikes is reduced
for larger coordination number.

\section{CONCLUSION}

Our investigations give evidence for the existence of the
entropy dominated phase. On the hypercubic triangulation  of the 4-torus
we find a
region which has well defined expectation values independent of the start
configuration even for unbounded action.

Calculations with irregular triangulations are still in their
infancy. While for limited fatness one finds only a shift of the
average curvature towards positive values the situation without cutoff
necessitates further simulations.

\end{document}